\documentclass{ws-ijmpa}

\begin{document}

\markboth{A. Bourque, C. Gale, and K. L. Haglin}
{$J/\Psi$ dissociation and Adler's theorem}

\catchline{}{}{}{}{}
%
%%%%%%%%%%%%%%%%%%%%%%%%%%%%%%%%%%%%%%%%%%%%%%%%%%%%%%%%%%%%%%%%%%%%

\title{$J/\Psi$ DISSOCIATION AND ADLER'S THEOREM}

\author{ALEX BOURQUE}

\address{Department of Physics, McGill University\\ 
3600 University Street, Montreal, QC, Canada H3A 2T8\\
bourquea@physics.mcgill.ca}

\author{CHARLES GALE}

\address{Department of Physics, McGill University\\ 
3600 University Street, Montreal, QC, Canada H3A 2T8\\
bourquea@physics.mcgill.ca}

\author{KEVIN L. HAGLIN}

\address{Department of Physics \& Astronomy, Saint Cloud State University
\\720 Fourth Avenue South, St. Cloud, MN 56301, USA\\
klhaglin@stcloudstate.edu}

\maketitle

\begin{abstract}

Effective Lagrangian models of charmonium have recently been used to
estimate dissociation cross sections with light hadrons.
Detailed study of the symmetry properties reveals possible
shortcomings relative to chiral symmetry.
We therefore propose a new Lagrangian and point out distinguishing 
features amongst the different approaches. Using the newly proposed Lagrangian, which 
exhibits ${\rm SU_L(4)\times\,SU_R(4)}$ symmetry and complies with
Adler's theorem, we find dissociation cross sections with pions that are 
reduced in an energy dependent way, with respect to cases where the theorem is
not fulfilled.

\end{abstract}

\ccode{PACS numbers: 11.25.Hf, 123.1K}

\section{Introduction}	
The theoretical study of matter under extreme conditions enjoys a wide range of 
application, from the physics of the early Universe, to that of relativistic 
nuclear collisions. The latter offer the tantalizing possibility of recreating 
in the laboratory the conditions that prevailed roughly a microsecond after the
Big Bang. The theory of the strong interaction, Quantum Chromodynamics (QCD),
predicts a phase transition from normal hadronic matter to a plasma of quarks
and gluons \cite{kl03}. To find a signature of this new state of matter
represents a task which has generated tremendous activity both in theory and in
experiment. Of the many signals put forward as probes of the quark-gluon plasma,
the suppression of the $J/\psi$ yield enjoys a popular status \cite{mat86}. 

Indeed,
since charmonium is predominantly produced in the early stage of the nuclear
collisions through hard processes, it acts as a probe for the subsequent stages.
The original idea was that the presence of  a quark-gluon plasma (QGP) will screen the
long-range confining force between $c$-$\bar c$, leading to the decoherence of
the pair \cite{mat86}. This suppression mechanism was later augmented by the
possibility of charmonium dissociation by hard gluons in a deconfined medium \cite{ks94}.
But, in order to identify the true nature of a possibly new phase, it appears necessary to
quantify the $J/\psi$-light hadron interaction.
This requirement on the $J/\psi$ cross-sections with light mesons
is not a trivial one to satisfy, as there are no direct experimental measurements. One
has to rely on theoretical calculations based, for example, on QCD sum-rules
\cite{dur03}, on quark-potential models \cite{mar95,won01}, on extented Nambu
and Jona-Lasinio models \cite{iva03,dea03}, or on effective ${\rm SU(4)}$
mesonic Lagrangians \cite{mat95,kh00,hag00,lin00,oh01}. These lead
typically to cross-sections from a few tenths of a millibarn to a few
millibarns near threshold \cite{dur03}. 

Specifically, for the
effective mesonic Lagrangians found in Refs~\refcite{hag00,lin00,oh01}, once form
factors have been folded-in  to account for short-range interactions, the
dissociation cross-section by pions reaches a few millibarns. But concerns have
been raised about using such models: (i) the ${\rm SU(4)}$ symmetry used to describe
the pseudoscalar and vector meson interactions is questionable as it is  
broken, (ii) the form factors accounting for the finite size of the mesons do
not proceed from the formalism as in other models \cite{dur03,won01}, and
(iii) the $J/\psi + \pi \rightarrow D^* + \bar D$ process does not vanish 
in the soft-pion
limit for non-degenerate vector meson masses as expected from Adler's theorem 
\cite{nav01}. In this contribution we will outline the consequence of implementing the soft-pion theorem 
(i.e., Adler's theorem) on the dissociation cross-section of 
$J/\psi + \pi \rightarrow D^* + \bar D$ by constructing a Lagrangian invariant under the global
${\rm SU_L(4) \times SU_R(4)}$ symmetry group.
 
%Full calculation details can be found in \cite{bou04}.

\section{Adler's theorem}

A spontaneously broken symmetry not only implies the existence of Goldstone 
bosons, but also constrains their low-energy behaviour. Here we will consider 
the transition amplitude for emitting one Goldstone boson (the proof can 
easily be extended to more than one). Following Weinberg \cite{wei95}, we first 
note that the current operator can create a Goldstone boson
\begin{equation}
\left<0|J^\mu(x)|B\right> = i\frac{Fq^\mu}{(2\pi)^3 2E} e^{-iq\cdot x},
\end{equation}
where F is the decay constant. We expect the matrix element $\left<\beta|J^\mu(x)|\alpha\right>$ to
then have a pole term. Indeed, in general, we can write
\begin{equation}
 \left<\alpha|J^\mu(0)|\beta\right> = N^\mu_{\beta\alpha} + i\frac{F q^\mu}{q^2}
 M^B_{\beta \alpha},
\end{equation} 
where $M^B_{\beta \alpha}$ is the desired transition amplitude for emitting one 
Goldstone boson, and $N^\mu_{\beta \alpha}$ is assumed to be the pole-free 
contribution. Applying the current conservation constraint, we then find that
\begin{equation}
M^B_{\beta \alpha} = \frac{i}{F}q_\mu N^\mu_{\beta \alpha} 
\end{equation}
and we see that as $q \rightarrow 0$ the RHS vanishes. Historically, this 
property was first studied by Adler \cite{adl65}, and rests on the assumption
that $N^\mu_{\beta\alpha}$ has no singularity as $ q\rightarrow 0$. This is not
valid in general, since a pole may arise through the insertion of the current 
operator on an external line \cite{sak69}. $M^B_{\beta \alpha}$ can then have 
a non-vanishing contribution in the soft limit due to emission of a soft 
Goldstone boson off an external line. But for chiral-symmetric and normal-parity mesonic 
Lagrangians such an insertion is impossible and we expect the theorem to hold. We 
will now explicitly check that decoupling occurs for the process
$\pi^+ + \rho^0 \rightarrow \pi^+ + \rho^0$ in two chiral-invariant
Lagrangians.

\subsection{Chiral model with $\mbox{\boldmath{$\pi$}}$, 
$\mbox{\boldmath{$\rho$}}$, and $\boldmath{\mbox{$a_1$}}$ mesons}
We first build the effective chiral Lagrangian as in Ref.~\refcite{mei88} yielding
\begin{eqnarray}
\mathcal{L} &=& \frac{1}{2}{\rm Tr}\left[\partial_\mu\phi\partial^{\mu}\phi\right] -\frac{1}{4}{\rm Tr}\left[F^V_{\mu\nu }F^{\mu\nu}_V +  F^A_{\mu\nu}F^{\mu\nu}_A \right] \nonumber  \\
&+&\frac{1}{2}m_V^2{\rm Tr}\left[V_\mu^2\right]
+\frac{1}{2}m_A^2{\rm Tr}\left[A_\mu^2\right] \nonumber \\
&+&\mathcal{L}_{V\phi\phi} + \mathcal{L}_{AV\phi} + \mathcal{L}_{VV\phi\phi} + \cdots\,,
\label{LVAP}
\end{eqnarray}
where $\phi$ is the pseudo-scalar field, $F_V$ and $F_A$ are the field strength tensors for the
vector and axial vector mesons respectively. The model has three parameters, namely the pion decay 
constant, $F_\pi$, 
the vector mass, $m_V$, and the universal coupling constant, $g$. The axial vector meson mass is
related to the vector mass in this model and thus is not an input parameter. 

\begin{figure}[!h]
\begin{center}
\includegraphics[scale=0.75]{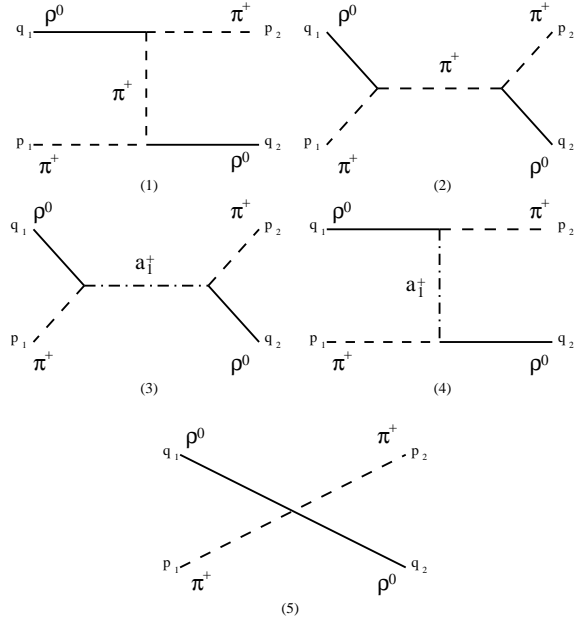}
\caption{$\rho^0 + \pi^+ \rightarrow \rho^0 + \pi^+$.}
\label{fig1}
\end{center}
\end{figure}

It is then seen that the process under consideration, namely $\pi^+(p_1)+\rho^0(q_1) \rightarrow 
\pi^+(p_2)+\rho^0(q_2)$, involves 5 tree-level diagrams for the given particle 
content: $\pi$- and $a_1$-mediated exchanges both in $s$ and $t$ channels, and 
through a 4-point interaction (Fig~\ref{fig1}).

Extracting the relevant
interaction terms results in the five off-shell amplitudes
\begin{eqnarray}
i\mathcal{M}^1_{\mu\nu} &=& i\Gamma^{1 \dagger}_{\mu}\frac{i}{s-m_\pi^2}
i\Gamma^{1}_\nu, \\
i\mathcal{M}^2_{\mu\nu} &=& i\Gamma^{1 \dagger}_{\mu}\frac{i}{t-m_\pi^2}
i\Gamma^{1}_\nu, \\
i\mathcal{M}^3_{\mu\nu} &=&
 i\Gamma^{2 \dagger}_{\mu\alpha}\frac{-i[g^{\alpha\beta}-
 (p_1+q_1)^\alpha(p_1+q_1)^\beta/m_A^2]}{s-m_A^2}i\Gamma^{2}_{\nu\beta},
\\
i\mathcal{M}^4_{\mu\nu}&=& i\Gamma^{2 \dagger}_{\mu\alpha}\frac{-i
[g^{\alpha\beta}-(q_2-p_1)^\alpha(q_2-p_1)^\beta/m_A^2]}{t-m_A^2}
i\Gamma^{2}_{\nu\beta},\\
i\mathcal{M}^5_{\mu\nu} &=& i\Gamma^{3}_{\mu\nu},
\end{eqnarray}
where the vertex functions are 
\begin{eqnarray}
\Gamma^1_{\mu} &=& -\frac{g}{\sqrt{2}}\left[k_\mu+ p_\mu -\frac{(1-Z^2)}
{m_V^2}\left(q\cdot k p_\mu-q\cdot p k_\mu\right)\right], \\
\Gamma^{2}_{\mu\nu} &=& -\frac{ig}{\sqrt{2}}\frac{(1-Z^2)^{\frac{1}{2}}}{m_V}
\left\{(m_A^2+q^2-k^2)g_{\mu\nu} -q_\mu q_\nu+k_\mu k_\nu\right \}, \\
\Gamma^{3}_{\mu\nu} &=& 2\left[\frac{g^2}{2}\frac{(1-Z^2)}{2m_V^2}\left\{-2p_1 
\cdot p_2 g_{\mu\nu} +
p_{1\mu}p_{2\nu} +p_{2\mu}p_{1\nu}\right\} + \frac{g^2}{2Z^2}g_{\mu\nu} \right],
\end{eqnarray}
for $\pi^+(p)+\rho^0(q) \rightarrow \pi^+(k)$, $\pi^+(p)+\rho^0(q) \rightarrow 
a_1^{+}(k)$, and the four-point interaction, respectively, and the shorthand 
$Z^2 = \left(1-\frac{g^2\tilde F_\pi^{\,2}}{4m_0^2}\right)$.

Having now the full amplitude for the given process, we wish to see if the 
pseudoscalar decoupling theorem holds. The presence of the contact term in the 
4-point interaction leads us to expect cancellations to occur as we let one
of the 
pions' 4-momentum go to zero. Stated differently, since the $\rho$-meson 
was introduced in a chirally symmetric way by adding its chiral partner 
(i.e. $a_1$), 
the transition amplitude relies on help from the $a_1$ in what
amounts to a delicate cancellation allowing the pion to
decouple.  To show 
this, we contract the amplitudes with the appropriate polarization vectors, 
and then let $p_2 \rightarrow 0$ (a similar proof can be shown to hold for $p_1 
\rightarrow 0$). First it is seen that the amplitudes involving the 
$\pi$-exchange 
go to zero because of transversality (i.e $\epsilon(q)\cdot q$).  For 
$a_1$-exchange in the $s$-channel we find
(the same result is true for the $t$-channel) 
\begin{eqnarray}
i\mathcal{M}_3 &=& \epsilon^{*\mu}(q_2)
\left[i\Gamma^{2\dagger}_{\mu\alpha}(p_2 \rightarrow 0) \frac{-i
\left[g^{\alpha\beta}-\frac{q_2^{\alpha}q_2^{\beta}}{m_A^2}\right]}
{m_V^2-m_A^2}i\Gamma^{2}_{\nu\beta}\right]\epsilon^{\nu}(q_1) \nonumber \\
&=& -\frac{ig^2}{2}\frac{1}{m_V^2}\epsilon^{*\mu}(q_2)\left[
m_A^2 g_{\mu\nu}-q_{1\mu}q_{1\nu} +\frac{q_1 \cdot q_2}{m_A^2}
q_{2\mu}q_{1\nu} + \frac{m_V^2}{m_A^2}q_{2\mu}q_{2\nu}\right]
\epsilon^{\nu}(q_1) \nonumber \\
&=& -\frac{ig^2}{2Z^2}\epsilon^*(q_2)\cdot \epsilon(q_1).
\end{eqnarray}
The last line comes about again due to the orthogonality condition. Finally, 
the 4-point interaction reads
\begin{equation}
i\mathcal{M}_5 = \frac{ig^2}{Z^2}\epsilon^*(q_2)\cdot \epsilon(q_1),
\end{equation}
and thus the {\it full amplitude} is shown to vanish as expected. Note the 
cancellation between the 4-point interaction and the $a_1$ channels. In summary,
even though the pions are not coupled through gradient coupling for all 
interaction terms, the net amplitude still vanishes due to intricate cancellations
amongst all the channels.

\subsection{Chiral model with $\boldmath{\mbox{$\pi$}}$ and 
$\boldmath{\mbox{$\rho$}}$ mesons}
In the previous section the axial mesons were introduced as the
chiral partners of the $\rho$ fields 
resulting in a linear realization of the symmetry \cite{gai69}. In the present 
case, since the desired chiral Lagrangian will involve 
only pseudoscalar and vector mesons, both will then have to transform 
non-linearly under the symmetry. This is done by gauging-away the axial 
mesons \cite{sch86} and yields
\begin{eqnarray}
\mathcal{L} &=& -\frac{1}{4}{\rm Tr}\left[F_{\mu\nu}(\rho)F^{\mu\nu}(\rho)\right] + \frac{1}{2}m_V^2{\rm Tr}[\rho_\mu^2] \nonumber \\
&+&\frac{F_\pi^{\,2}}{4}(1+k){\rm Tr}\left[\partial_{\mu}\zeta\partial^{\mu}\zeta\right]+ i\frac{F_\pi^{\,2}\,g_{V\phi\phi}}{2}{\rm Tr}\left[\rho^\mu\left(\partial_\mu\zeta\zeta^\dagger + \partial_\mu\zeta^\dagger\zeta\right)\right] \nonumber \\
&+&\frac{F_\pi^{\,2}}{4}(1-k){\rm Tr}\left[\zeta^\dagger\partial^\mu\zeta^\dagger\zeta\partial_\mu\zeta\right],
\label{LVP1}
\end{eqnarray}
where the input parameters are now $m_V^2$, $g_{V\phi\phi}$, $F_\pi$, and $k$, 
and the field strength tensor is 
$F_{\mu\nu}(\rho) = \partial_\mu \rho_\nu - \partial_\nu \rho_\mu -ig[\rho_\mu,
\rho_\nu]$.

In this model there are only two diagrams, namely $s\,$- and $t\,$-channels of 
pion exchange. The relevant interaction Lagrangian is
\begin{equation}
\mathcal{L}_{\rho\pi\pi} = -i\frac{g_{\rho\pi\pi}}{2}{\rm Tr}
\left[\rho^\mu[\phi,\partial_\mu \phi]\right],
\end{equation}
and the extracted vertex for the s- and t- channels is
\begin{eqnarray}
\Gamma_\mu^{'1} = -\frac{g_{\rho\pi\pi}}{\sqrt 2}\left(p_\mu+k_\mu\right), \\
\end{eqnarray}
with $k=p+q$ and $k=p-q$, respectively. The two amplitudes are then
\begin{eqnarray}
i\mathcal{M}^1_{\mu\nu} = i\Gamma^{'1 \dagger}_{\mu}\frac{i}{s-m_\pi^2}i\Gamma^{'1}_\nu\, , \\
i\mathcal{M}^2_{\mu\nu} = i\Gamma^{'1 \dagger}_{\mu}\frac{i}{t-m_\pi^2}i\Gamma^{'1}_\nu\,.
\end{eqnarray}
We immediately see that the soft pion theorem holds separately for each 
amplitude when the proper polarization vectors are contracted with the 
amplitudes.

\section{Application to the $J/\Psi$ dissociation}
\begin{figure}[!h]
\begin{center}
\includegraphics[scale=0.75]{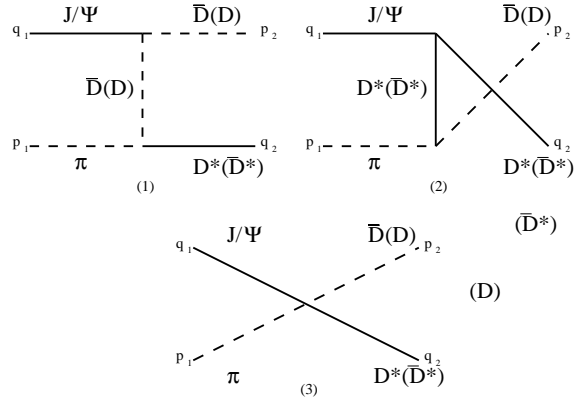}
\caption{$J/
\psi + \pi \rightarrow D^{*}(\bar D^{*}) + \bar D(D)$.}
\label{fig2}
\end{center}
\end{figure}

We first start by noting that all the effective Lagrangian models found in the literature 
\cite{hag00,lin00,oh01} are invariant under ${\rm SU_V(4)}$ (in the degenerate mass 
limit), but not under the extended chriral symmetry, i.e., ${\rm SU_L(4) \times SU_R(4)}$, and therefore
we expect the pions will not decouple in soft-momentum limit. Specifically, here we will compare the 
$J/\psi + \pi \rightarrow D^* + \bar D$ cross-section as calculated within the ${\rm SU_V(4)}$--only models found 
in Refs~\refcite{hag00,lin00,oh01} and with the Lagrangian of Eq.~(\ref{LVP1}) extended to the ${\rm
SU_L(4) \times SU_R(4)}$
symmetry group.
For the ${\rm SU_V(4)}$ models, three diagrams need to be considered 
(Fig.~\ref{fig2}), while for the extended chiral model proposed here, the contact interaction 
is absent.  It can be explicitly shown that the pseudoscalars decouple in the soft momentum limit 
for that latter case \cite{bou04}.

\begin{figure}[!h]
\begin{center}
\includegraphics[scale=0.50]{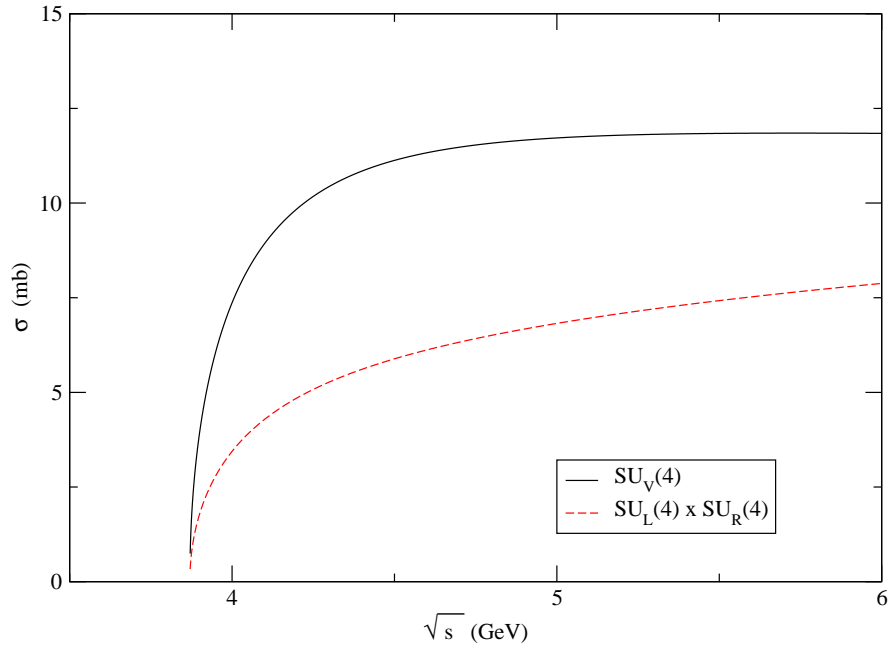}
\caption{Isospin-averaged cross-section for $J/\psi + \pi \rightarrow (D^* + 
\bar D) + (\bar D^* + D) $.}
\label{fig3}
\end{center}
\end{figure}

The differential isospin-averaged cross-section 
is given by
\begin{equation}
\frac{d\sigma}{dt} = \frac{1}{128\pi s p^2_{1}} 
\mathcal{M}^{\mu\nu}\mathcal{M}^{\alpha\beta}\left[g_{\mu\alpha}-
\frac{q_{2\mu}q_{2\alpha}}{m^2_{D^*}}\right]\left[g_{\nu\beta}-\frac{q_{1\nu}q_{1\beta}}
{m^2_{J/\psi}}\right],
\end{equation} 
where the appropriate model-dependent squared amplitude is used, an
isospin factor 
of two has been included and the centre of mass momentum is 
\begin{equation}
p^2_{1} = \frac{1}{4s}\lambda(s,m^2_\pi,m^2_{J/\psi})
\end{equation}
and the triangle function is $\lambda(x,y,z) = x^2 -2x(y+z)+(y-z)^2$.
Integrating over the kinematical range defined by
\begin{eqnarray}
t_\pm &=& m_\pi^2 + m_{D^*}^2 - \frac{1}{2s}(s+m_\pi^2 - m_{J/\psi}^2)(s+m_{D^*}^2 
- m_{D}^2) \nonumber \\
&\pm& \frac{1}{2s}\lambda^{1/2}(s,m_\pi^2,m_{J/\psi}^2)\lambda^{1/2}
(s,m_D^2,m_{D^*}^2)
\end{eqnarray}
gives the total cross-section. Fixing the relevant parameters \cite{bou04} and carrying this to completion 
for the two models yields Fig.~\ref{fig3}. 
We see an energy-dependent reduction in the cross sections across 
the relevant domain and to quote a specific number we note
that at $\sqrt s = 5$ GeV the cross-section is reduced by 
about 40\% going from the ${\rm SU_V(4)}$ model to the ${\rm SU_L(4) \times SU_R(4)}$ model.   

\section{Conclusion}
Modeling low-energy hadron physics is particularly challenging
when there is limited experimental input available
for constraints.  This is the current situation in the 
charm sector as the only measurement relevant for fixing
coupling constants in the model is the
decay width for $D^{*}\rightarrow\/D\pi$~\cite{cleo}.
We have therefore invoked symmetries and general theorems.
In particular, we have checked for full ${\rm SU_L(4)\times\/SU_R(4)}$
symmetry and the appropriate limit to test for compliance with Adler's 
theorem. We found that none of the published models can 
do this, and we therefore proposed a new effective Lagrangian---the
first one which does encode complete four-flavour chiral symmetry and Adler's theorem.
%Rather than carry out a full model calculation at this
%stage, complete with hadronic form factors,
Our interest here has been solely to quantify the effect of these.  A complete calculation including form factors and a longer list of reactions is a topic for a separate study.

Since Adler's theorem is relevant at low-energy, the
near-threshold cross sections are expected to be affected
the most.   We found the cross section for $J/\psi+\pi\rightarrow
(D^{*}+\bar{D}) + h.c.\,$ to be reduced as compared to   
a choice of Lagrangian which does not encode the full flavour 
chiral symmetry and does not obey Adler's theorem.  The reduction
is energy dependent, but seems to be a few tens of percents
from threshold to $\sqrt{s}$ = 5 GeV. In a full calculation the size of this reduction 
might not persist when one takes into account not only form factors, but also 
abnormal parity interactions and symmetry breaking effects (e.g. pseudoscalar masses 
and non-degenerate vector mass spectrum). 

Abnormal parity interactions may play an important role near threshold. Indeed, it was shown that Adler's theorem breaks down if a soft Goldstone boson can be inserted on an external line. This is expected to happen for abnormal parity Lagrangians where a $VV\phi$ vertex exists. The abnormal parity contribution to the $J/\psi +\pi$ amplitude will then not vanish in the soft limit. The problem in including these interactions lies again in the lack of experimental data to fix the coupling strengths. 

Symmetry breaking effects are also expected to be important since the underlying ${\rm SU_L(4)\times
SU_R(4)}$ is broken. 

It will also be important in the
future to take this formalism to completion by implementing
covariant hadronic form factors computed within the same
effective Lagrangian or perhaps other approaches.  Ultimately, the outlook
for this line of study is to estimate the dissociation cross
sections with all the light hadrons, with finite size effects
incorporated, and then to input the results into a
dynamical model for heavy ion reactions to finally address
the question of $J/\psi$ survivability in the hadronic
phase (primarily mesonic matter).
For then, one would have a more complete understanding of the $J/\psi$ yield 
and therefore know what it implies about QGP formation.

\section*{Acknowledgments}

A.B. thanks S. Turbide for helpful discussions, and A.B. and C.G. thank E. S. Swanson 
for  a useful visit. 
This work was supported in part by the Natural Sciences and Engineering 
Research Council of Canada, in part by the Fonds Nature et Technologies
of Quebec, and in part by the National Science Foundation under
grant number PHY-0098760.

%\section{References}

%\begin{thebibliography}{000} %for 3 digits
%\begin{thebibliography}{00}  %for 2 digits


\begin{thebibliography}{99}    %for 1 digit
\bibitem{kl03} F. Karsch and E. Laermann, {\it Quark-Gluon Plasma 3}, 
eds Rudolph C. Hwa and  Xin-Nian Wang (World Scientific, Singapore, 2004).
\bibitem{mat86} T. Matsui and H. Satz, {\it Phys. Lett.} {\bf B178}, 416 (1986).
\bibitem{ks94} D. Kharzeev and H. Satz, {\it Phys. Lett.}  {\bf B334}, 155 (1994).
%\bibitem{na38} M. C. Abreu {\it et al.}, {\it Phys. Lett.} {\bf B449}, 128 (1999).
%\bibitem{na50} M. C. Abreu {\it et al.}, {\it Phys. Lett.} {\bf B477}, 28 (2000).
%\bibitem{arm98} N. Armesto and A. Capella, {\it Phys. Lett.} {\bf B430}, 23 (1998).
%\bibitem{cap02} A. Capella, A. B. Kaidalov, and D. Sousa, {\it Phys. Rev.} {\bf C65}, 054908 (2002).
\bibitem{dur03} D. Kharzeev, H. Satz, A. Syamtomov, and G. Zinovjev, {\it Phys. Lett.} {\bf B389},
595 (1996); F.~O.~Duraes, H.~C.~Kim, S.~H.~Lee, F.~S.~Navarra, and 
M.~Nielsen, {\it Phys. Rev.}  {\bf C68}, 035208 (2003).
\bibitem{mar95} K. Martins, D. Blaschke, and E. Quack,{\it Phys. Rev.} {\bf 51}, 
2723 (1995).
\bibitem{won01} C.-Y. Wong, E.S. Swanson, and T. Barnes,{\it Phys. Rev.}
{\bf C65}, 014903 (2001). 
\bibitem{iva03} M.A. Ivanov, J. G. K\"orner, and P. Santorelli, {\it Phys. Rev.}
{\bf D70}, 014005 (2004). 
\bibitem{dea03} A. Deandrea, G. Nardulli, and A. D. Polosa,{\it Phys. Rev.} {\bf
D68},034002 (2003). 
\bibitem{mat95} S. G. Matinyan and B. M\"uller,{\it Phys. Rev.} {\bf C51}, 2723 
(1995).
\bibitem{kh00} K. L. Haglin,{\it Phys. Rev.} {\bf C61}, 031902 (2000).
\bibitem{hag00} K. L. Haglin and C. Gale, {\it Phys. Rev.} {\bf C63}, 065201 (2001).
\bibitem{lin00} Z. Lin, C.M. Ko, and B. Zhang,{\it Phys. Rev.} {\bf C61}, 024904 
(2000).
\bibitem{oh01} Y. Oh, T. Song, and S. Houng Lee, {\it Phys. Rev.} {\bf C63}, 
034901 (2001).
\bibitem{nav01} F. S. Navarra, M. Nielson, and M. R. Robilotta, {\it Phys. Rev.}
{\bf C64}, 021901 (2001).
\bibitem{wei95} S. Weinberg, {\it The Quantum Theory of Fields}, Vol. 2 (Cambridge University Press, 1995). 
\bibitem{adl65} S. Adler, {\it Phys. Rev.} {\bf 137}, B1022 (1965).
\bibitem{sak69} J.J. Sakurai, {\it Currents and Mesons} (University of 
Chicago Press, Chicago, 1969).
\bibitem{mei88} U. Meissner, {\it Phys. Rept.} {\bf 161}, 215 (1988). 
\bibitem{gai69} S. Gasiorowicz and  D. A. Geffen, {\it Rev. Mod. Phys.} {\bf 41}, 531 
(1969).
\bibitem{sch86} J. Schechter,{\it Phys. Rev.} {\bf D34}, 868 (1986).
\bibitem{bou04} A. Bourque, C. Gale, H. L. Haglin, {\it Phys. Rev.} {\bf C70}, 055203 (2004)
\bibitem{cleo} A. Anastassov {\it et al.}, {\it Phys. Rev.} {\bf D65}, 032003 (2002)



\end{thebibliography}
\end{document}